\documentclass{article}
\usepackage[a4paper, total={6.5in, 9in}]{geometry}
\usepackage[english]{babel}
\usepackage{graphics}% Include figure files
\usepackage{dcolumn}% Align table columns on the decimal point
\usepackage{bm}% bold math
\usepackage{hyperref}% add hypertext capabilities
\hypersetup{
	colorlinks=true,
	linkcolor=blue,
	filecolor=blue,      
	urlcolor=blue,
	pdftitle={Overleaf Example},
	pdfpagemode=FullScreen,
}
\usepackage{mathtools, amsfonts, amssymb}
\usepackage{amsthm}
\usepackage{braket}
\usepackage{tikz}
\usepackage{tikz-cd}
\usepackage{tikzit}
\tikzstyle{every picture}=[baseline=-0.25em,]
\usepackage[all,arrow,matrix]{xy} 
\newtheorem{definition}{Definition}

\newtheorem{rem}{Remark}

\ifpdf
  \usepackage{underscore}         % Only needed if you use pdflatex.
  \usepackage[T1]{fontenc}        % Recommended with pdflatex
\else
  \usepackage{breakurl}           % Not needed if you use pdflatex only.
\fi

\title{The 2-Category of Topological Quantum Computation}
\author{Fatimah Rita Ahmadi\\Department of Mathematics, Imperial College London}

\begin{document}
\maketitle
\begin{abstract}
Unitary Ribbon Fusion Categories (URFC) formalize anyonic theories. It has been widely assumed that the same category formalizes a topological quantum computing model. However, in \cite{ahmadi2022top}, we addressed and resolved this confusion and demonstrated while the former could be any fusion category, the latter is always a subcategory of \textbf{Hilb}. In this paper, we argue that a categorical formalism that captures and unifies both anyonic theories~(the Hardware of quantum computing) and a model of topological quantum computing is a braided (fusion) 2-category. In this 2-category, 0-morphisms describe anyonic types and Hom-categories describe different models of quantum computing. This picture provides an insightful perspective on superselection rules. It presents furthermore a clear distinction between fusion of anyons versus tensor products as defined in linear algebra, between vector spaces of 1-morphisms. The former represents a monoidal product and sum between 0-morphisms and the latter a tensor product and direct sum between 1-morphisms. 
\end{abstract}
\section{Introduction}
Topological quantum computation, proposed by Alexei Kitaev and Michael Freedman \cite{kitaev2003, freedman2002}, is an intrinsically robust model of quantum computation that resists errors. In this framework, qubits are encoded in the collective states of various configurations of anyons. Unlike bosonic and fermionic particles in three dimensions, whose statistics are governed by permutation groups, anyons follow braid group statistics. Quantum gates correspond to exchanges of anyons and are realized through unitary representations of the braid group. In systems with gapped and local Hamiltonians, and under suitable conditions, the computation is robust against local perturbations, as local errors cannot induce transformations or exchanges that would implement unintended gates \cite{sarma2006}.

The mathematical formalism for anyonic theories is developed using category theory \cite{wang2010}. This shift from set theory to category theory arises from the unique properties of anyons and, in most systems, from the presence of topological background fields. Restricting our attention to point-like anyons in two-dimensional topologically ordered systems, we note that anyons possess fractional charges and spins. As a result, assigning a Hilbert space to a single anyon using the conventional formalism is not well-defined. Instead, qubits are encoded in the joint states of multiple anyons. In other words, the operations between anyonic states constitute the algebraic structure of physical operations, which themselves form well-defined Hilbert spaces.

Another motivation for the shift in language is historical. Inspired by categorical conformal field theories developed by Singer, Michael Atiyah proposed a categorical framework for topological quantum field theories \cite{atiyah1988}. He suggested that such a theory can be formulated as a functor from the category of cobordisms to the category of vector spaces. Given that anyonic statistics are typically observed in systems with topological background fields, the use of category theory becomes a natural and almost inevitable development.

In the categorical formulation of an anyonic theory, objects correspond to anyon types, hom-sets to processes between different configurations, the monoidal product of objects to fusion rules or mutual statistics, and braidings to the braid-like exchanges of anyons. This same categorical framework is widely believed to provide a suitable foundation for topological quantum computation. Note that the computational spaces involved in such models are Hilbert spaces corresponding to operations between different configurations of anyonic states. Within this framework, qubits are represented by states within the hom-sets of the category \cite{wang2010}.

In previous work \cite{ahmadi2022top}, we argued that the prevailing approach combines two conceptually distinct categories into a single framework: one for anyonic theory and another for the model of topological quantum computation. The first category describes the physical theory—it includes objects (anyon types), their monoidal products (fusion rules), associators, and braiding morphisms. The second category is concerned with computation: it involves Hilbert spaces of operations (i.e., hom-sets), fusion matrices, and braiding matrices. The former constrains the physical properties of the system, while the latter defines a computational model. We demonstrated that the category associated with topological quantum computation is a subcategory of the category of Hilbert spaces.

These observations lead us to a central question: how can one categorically describe both anyonic theory and topological quantum computation in a unified framework? This paper proposes that the unifying structure is a braided fusion 2-category, where:
\begin{itemize}
\item 0-morphisms correspond to anyon types,
\item 1-morphisms to vector spaces, and
\item 2-morphisms to fusion and braiding matrices.
\end{itemize}
\begin{rem}
We assume the reader has an elementary knowledge of category theory at the level of Leinster's book \textit{Basic Category Theory}~\cite{leinster2014}. For a review on bicategories, Check~\cite{leinster1998basic}. 
\end{rem}
\begin{rem}
By bicategories, we mean strict 2-categories (i.e. strictness on the composition of 1-morphisms). To review a complete definition of monoidal 2-categories, see~\cite{ahmadi2020monoidal} and for braided monoidal 2-categories, see~\cite{baez1996higher}. Note that Baez and Neuchl's definition of monoidal 2-categories is semistrict, whereas~\cite{ahmadi2020monoidal} does not assume the strictness of any monoidal data. Here, we work with strict braided monoidal 2-categories. 
\end{rem}
\begin{rem}
To further learn the physics of systems with topological orders, See~\cite{simon2023topological}. 
\end{rem}

\section{A prologue on the problem}
The initial motivation for this paper stems from the following observation: the categorical description of topological quantum computation currently employs two distinct types of tensor and direct products~\cite{rowell2018mathematics, wang2010}. The first type involves tensor and direct products between anyons, describing how anyons interact. These operations capture, for example, the creation of electric and magnetic charges via loops in the Toric code~\cite{kitaev2003}, or the fusion and exchange processes of anyons in fractional quantum Hall systems~\cite{simon2023topological}.

The second type involves tensor and direct products between vector spaces that encode morphisms between different configurations of anyons. These structures are studied within the framework of linear algebra.

Currently, the literature tends to treat these two types of product structures interchangeably. However, the issue is not merely notational. When asked to precisely distinguish between these two forms, the answers typically remain unsatisfactory. The problem lies in the fact that two separate monoidal and biproduct structures are at play—one among anyons and one among vector spaces—but they are often conflated or treated informally. It is mistakenly assumed that they share the same categorical nature or behave identically.

A further issue concerns the superselection rule, which is typically imposed as an external axiom rather than integrated into the categorical framework. This becomes critical when considering coherent state constructions. Specifically, the sum of two vectors from different distinguished fusion spaces—such as $\ket{\psi_1} \in V_{xy}^z$ and $\ket{\psi_2} \in V_{xy}^k$—does not yield a coherent physical state. Our aim is to embed the superselection rule into the categorical formalism itself, so that it emerges naturally from the structure, rather than being externally enforced.
\section{The Category of an Anyonic Theory}
As previously mentioned, anyonic theories describe the hardware underpinning models of topological quantum computation. Any such model requires a collection of anyonic excitations whose statistical behaviour is governed by braid groups. Physically, anyons are quasiparticle excitations in two-dimensional topologically ordered materials and exhibit properties distinct from bosons and fermions in three-dimensional space—for instance, they may carry fractional charges or possess fractional spin~\cite{sarma2006, simon2023topological}.

Another crucial property of anyons is their non-trivial fusion behaviour. In contrast to bosons and fermions, which fuse uniquely to produce a well-defined particle with specific spin and charge, the fusion of anyons can yield multiple possible outcomes. When this is the case, they are referred to as non-Abelian anyons. These are particularly useful for topological quantum computation, as their fusion spaces are multidimensional, enabling the encoding and manipulation of quantum information. Various experimental proposals aim to realize such non-Abelian anyons, with Ising anyons—also known as Majorana fermions—being among the most prominent candidates~\cite{alicea2012new}. The fusion of Ising anyons yields a two-dimensional configuration space, which allows for qubit encoding. However, this model is not universal for quantum computation, as it only supports the implementation of Clifford gates. Nonetheless, it offers intrinsic topological protection against some local errors. A recent development by Microsoft, the Majorana 1 chip, is based on a single-qubit Majorana fermion~\cite{aasen2025roadmap}. The specifics of these experimental setups and proposals lie beyond the scope of this paper.

Due to properties such as fractional spin and charge, a conventional Hilbert space description of anyonic systems is not straightforward, particularly in light of the standard formalism developed for bosons and fermions in statistical physics. A categorical description, on the other hand, treats anyons as objects in a category, abstracting away from their detailed physical characteristics. Morphisms in this framework correspond to physical processes—such as creation and annihilation operators—between anyonic states. The appropriate categorical structure for capturing the features of anyonic systems is a unitary ribbon fusion category~\cite{wang2010, rowell2018mathematics}. These are semisimple categories with finitely many simple objects; all other objects are constructed as tensor products or direct sums of these simple objects. The tensor product structure requires the category to be monoidal, with associativity encoded by the pentagon equation. The exchange statistics of anyons are described by a braiding, which satisfies the hexagon equation. Additional properties, such as topological spin, are encoded by a ribbon structure or twist. To ensure the physicality of the theory, the category must be unitary, meaning that all solutions to the pentagon and hexagon equations are unitary transformations.

We do not provide full definitions of these structures here. (A more comprehensive account, including a review of unitary ribbon fusion categories, may be included in a future version of this work.)

\begin{definition}[\textbf{Anyonic Category}]
An anyonic category is unitary braided fusion category, such that it includes the following structures and properties. 
\begin{itemize}
    \item \textbf{Objects:} Objects are simple and represent anyon types. Each object has a \textbf{dual object} representing its anti-particle. This is captured by \textbf{rigidity} of the category. The object set includes a unit object, $I$. We represent simple objects with $X_i$, and $X_i \cong I$ where $i=0$. Rigidity now means for every index, $i$, there exist an object with index $j$ such that $X_i \times X_j \cong X_0$. We represent a dual object with a star sign, meaning, $X_i \times (X_i)^* \cong (X_i)^* \times X_i \cong X_0$. 
    \item \textbf{Morphism:} Morphisms belong to hom-sets indexed by three labels, $hom(X_i \times X_j, X_k) = V_{ij}^k$. Because objects are simple, hom-set between simple objects only include zero morphism, $hom(X_i, X_j)\cong 0$. The non-trivial morphisms belong to hom-sets called fusion spaces, $V_{ij}^k$. Since the category has adjoints, for each fusion space, we have the isomorphic splitting space, $hom(X_k, X_i \times X_j ) \cong V_{k}^{ij}$. The hom-sets are furthermore vector spaces. A 1-morphism $X_i \times X_j \longrightarrow X_k$ is called an annihilation operator and a 1-morphism $X_k \longrightarrow X_i \times X_j$ is called a creation operator. 
    \item \textbf{Monoidal:} The category is monoidal. There exist a monoidal product between each pair of objects $X_i \otimes X_j$ with an isomorphim $F_{ijk}$ as an associator between three objects, 
    \[
    (X_i \times X_j)\times X_k \overset{F_{ijk}}{\cong}  X_i \times ( X_j\times X_k)
    \]
    such that it satisfies the pentagonal equation~\ref{fig:pentag1}. The solutions of the pentagonal equation are unitary matrices. 
    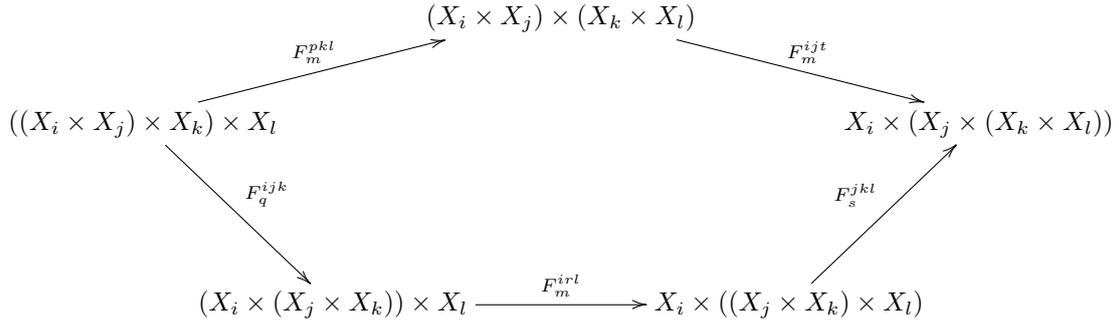
\begin{figure}
        \centering
\begin{equation*}
\begin{xy}
(-55,0)*+{((X_i \times X_j) \times X_k)\times X_l}="1",
(0,14)*+{(X_i \times X_j) \times (X_k \times X_l)}="2",
(55,0)*+{X_i \times (X_j \times (X_k \times X_l))}="3",
(-30,-24)*+{(X_i \times (X_j \times X_k))\times X_l}="4",
(30,-24)*+{X_i \times ((X_j \times X_k )\times X_l)}="5",
\ar^-{F_{m}^{pkl}} (-4,0)+"1"; (-4,0)+"2"
\ar^-{F_{m}^{ijt}} (4,0)+"2"; (4,0)+"3"
\ar^{F_{q}^{ijk}} "1";"4"
\ar^{F_{m}^{irl}} "4";"5"
\ar^{F_{s}^{jkl}} "5";"3"
\end{xy}
\end{equation*}
        \caption{Pentagonal equations with simple objects. }
        \label{fig:pentag1}
    \end{figure}
\item \textbf{Additive: } The category is additive, meaning between each pair of objects we have a well-defined addition. Particularly, since the tensor product between simple objects creates non-simple objects, the outcome can be written as an addition of anyon types. The sum is finite. 
\[
X_i \times X_j \cong X_k + X_l + \dots + X_m
\]
The nature of addition is a biproduct between objects.  If you compose an object with its dual, the unit object is definitely an outcome. 
\[
X_i \times (X_i)^* = I + X_k + \dots + X_k
\]
\item \textbf{Braided: } The category is braided; meaning for every two objects, we have an isomorphism which swaps two anyons, such that they satisfy hexagonal equations~\ref{hex} and the solutions are unitary. 
\[
X_i \times X_j \overset{R_{ij}}{\cong} X_j \times X_i 
\]
\end{itemize}
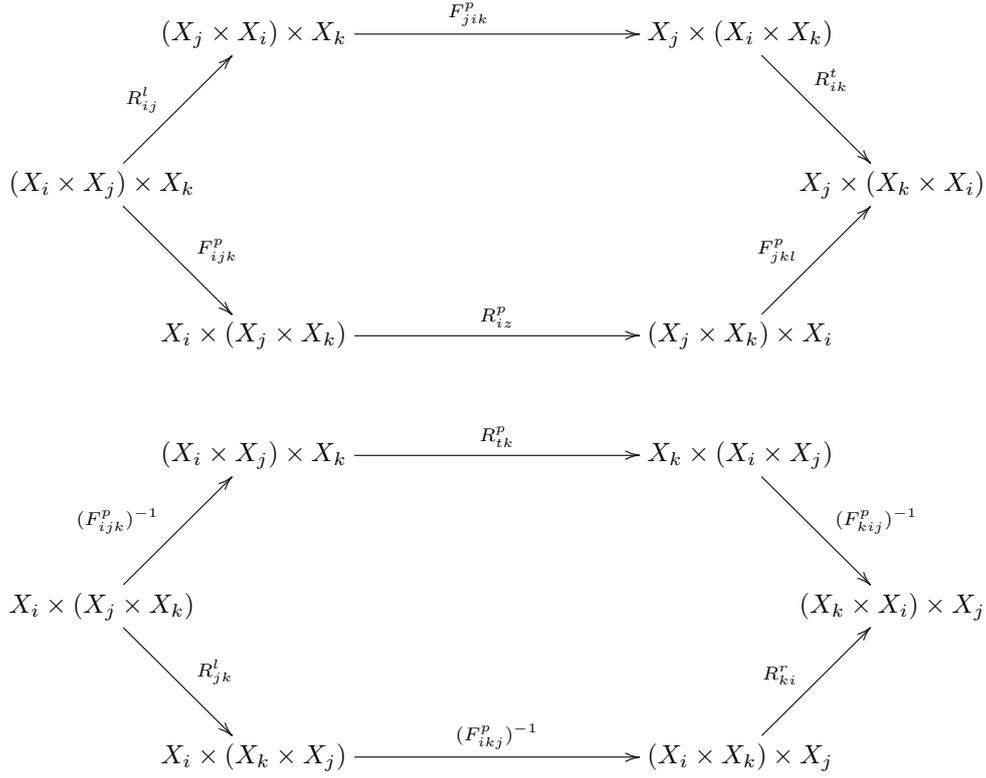
\begin{figure}[!ht]
\centering
\begin{equation*}
\begin{xy}
(-52,0)*+{(X_i \times X_j)\times X_k}="1",
(-32,20)*+{(X_j \times X_i) \times X_k}="2",
(32,20)*+{X_j \times(X_i \times X_k)}="3",
(52,0)*+{X_j \times(X_k \times X_i)}="4",
(-32,-20)*+{X_i \times(X_j \times X_k)}="5",
(32,-20)*+{(X_j \times X_k) \times X_i}="6",
\ar^-(0.4){F_{jik}^p} "2";"3"
\ar^-(0.6){F_{ijk}^p} "1";"5"
\ar^-{R_{iz}^p} "5";"6"
\ar^-(0.4){F_{jkl}^p} "6";"4"
\ar^-(0.4){R_{ij}^l} "1";"2"
\ar^-(0.4){R_{ik}^t} "3";"4"
\end{xy}
\bigskip
\end{equation*}
\begin{equation*}
\begin{xy}
(-52,0)*+{X_i \times(X_j \times X_k)}="1",
(-32,20)*+{(X_i \times X_j) \times X_k}="2",
(32,20)*+{X_k \times(X_i \times X_j)}="3",
(52,0)*+{(X_k \times X_i) \times X_j}="4",
(-32,-20)*+{X_i \times(X_k \times X_j)}="5",
(32,-20)*+{(X_i \times X_k) \times X_j}="6",
\ar^-(0.4){(F_{ijk}^p)^{-1}} "1";"2"
\ar^-{R_{tk}^p} "2";"3"
\ar^-(0.6){(F_{kij}^p)^{-1}} "3";"4"
\ar^-(0.6){R_{jk}^l} "1";"5"
\ar^-{(F_{ikj}^p)^{-1}} "5";"6"
\ar^-(0.4){R_{ki}^r} "6";"4"
\end{xy}
\end{equation*}
\caption{Hexagonal equations with simple objects.}\label{hex}
\end{figure}
\end{definition}
\begin{rem}
The monoidal product, denoted by $\times$, representing the fusion or composition of anyons, is fundamentally different from the tensor product $\otimes$ or the Cartesian product commonly encountered in linear algebra. In fact, much of the richness of anyonic theories arises from this distinction. Similarly, the addition of objects in the category corresponds to a biproduct, whose existence depends on the categorical properties of limits and colimits, rather than being imposed as an external structure. It is important to emphasize that this additive operation, typically denoted by $+$, is not the familiar direct sum $\oplus$ from linear algebra.
\end{rem}

\section{Topological Quantum Computation}
Although unitary braided fusion categories fully capture the properties and structures expected of an anyonic theory, they are often conflated with the category underlying topological quantum computation (TQC). In other words, it is frequently assumed that the computational category is identical to the category of anyons. However, as argued earlier, the TQC category is more accurately viewed as a \emph{subcategory of the category of Hilbert spaces}, denoted by~\(\mathbf{Hilb}\). Within this framework, the matrix representations of the \(F\)- and \(R\)-symbols acquire direct physical and computational interpretations.

Let us now review the essential elements required for TQC. A well-defined anyonic theory provides the necessary data: a finite set of anyon types, fusion rules (which describe the decomposition of the monoidal product of anyon types), and consistent solutions to the \emph{pentagon} and \emph{hexagon equations}.

The \emph{computational space} is not the space of anyons themselves, but rather the \emph{fusion space}~\(V^k_{ij}\). Physically, qubits are encoded in the degrees of freedom associated with different possible fusion outcomes of anyons. In this setting, the fusion space \(V^k_{ij}\) serves as a computational space if the fusion \((X_i \otimes X_j) \otimes X_k\) admits more than one possible outcome—either due to multiplicity \(N_{ij}^k = 2\), or because multiple fusion channels \(k\) are available.

Given this picture, the solutions to the \emph{pentagon equation}—the so-called \(F\)-matrices—are realized as matrix-valued linear isomorphisms between associatively parenthesized fusion spaces:
\[
V^l_{(ij)k} \cong V^l_{i(jk)}.
\]
Since the building blocks are the fusion spaces \(V^k_{ij}\), more complex spaces like \(V^l_{(ij)k}\) decompose as:
\[
V^l_{(ij)k} \cong \bigoplus_e V^e_{ij} \otimes V^l_{ek}.
\]
Similarly, the solutions to the \emph{hexagon equation}—the \(R\)-matrices—are linear isomorphisms between fusion spaces corresponding to the exchange of anyons:
\[
V^k_{ij} \cong V^k_{ji}.
\]
These matrix representations provide the tools for implementing quantum gates in TQC, since braiding operations correspond to unitary transformations on the fusion spaces in which qubits are encoded.
\begin{definition}
A TQC category consists of the following:
\begin{itemize}
\item \textbf{Objects} fusion spaces, the direct sums and tensor products of fusion spaces $V_{ij}^k$. (This category is furthermore semi-simple: simple objects are fusion spaces, and any other space is a direct sum of fusion spaces.) 
\begin{equation}
	V_{i_1..i_n}^j \cong \bigoplus_{k_1...k_{n-1}} V_{i_1i_2}^{k_1} \otimes  V_{k_1i_3}^{k_2}\otimes  ... \otimes V_{k_{n-1}i_n}^{j}
\end{equation}
\item \textbf{Morphisms} $R$ matrices, $F$ matrices, and compositions and tensor products thereof.
\item \textbf{Pentagonal and Hexagonal Equations} $F$-matrices between fusion spaces satisfy the pentagonal equation~\ref{fig:penta2} and $R$-matrices satisfy the hexagonal equations~\ref{fig:hex1}(This version is given in the Appendix E of \cite{kitaev2006}). 
\end{itemize}
\end{definition}
\begin{figure}
    \centering
\begin{equation*}
\begin{xy}
(-55,0)*+{\bigoplus_{p,q}V_{p}^{xy}\otimes V_{q}^{pz}\otimes V_{u}^{qw}}="1",
(0,14)*+{\bigoplus_{p,t}V_{p}^{xy}\otimes V_{u}^{pt}\otimes V_{t}^{zw}}="2",
(55,0)*+{\bigoplus_{s,t}V_{u}^{xs}\otimes V_{s}^{yt}\otimes V_{t}^{zw}}="3",
(-30,-24)*+{\bigoplus_{q,r}V_{q}^{xr}\otimes V_{r}^{yz}\otimes V_{u}^{qw}}="4",
(30,-24)*+{\bigoplus_{r,s} V_{u}^{xs}\otimes V_{r}^{yz}\otimes V_{s}^{rw}}="5",
\ar^-{F_{u}^{pzw}} (-4,0)+"1"; (-4,0)+"2"
\ar^-{F_{u}^{xyt}} (4,0)+"2"; (4,0)+"3"
\ar^{F_{q}^{xyz}} "1";"4"
\ar^{F_{u}^{xrw}} "4";"5"
\ar^{F_{s}^{yzw}} "5";"3"
\end{xy}
\end{equation*}
\caption{Pentagonal equation with fusion spaces.}\label{fig:penta2}
\end{figure}
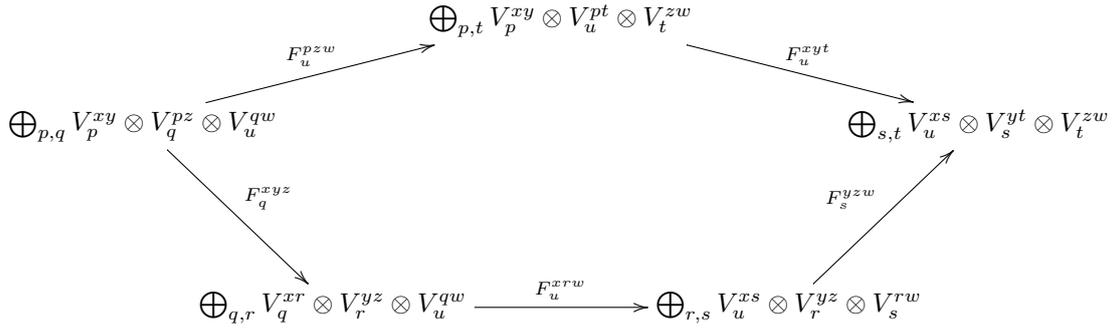
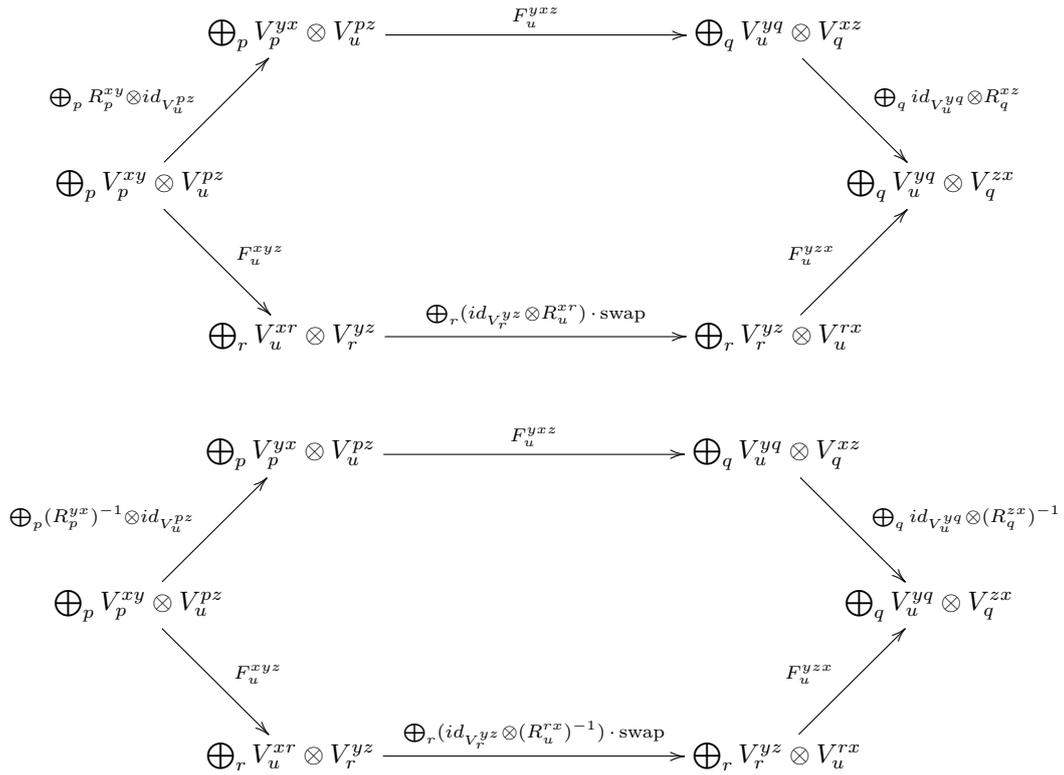
\begin{figure}[!ht]
    \centering
\begin{equation*}
\begin{xy}
(-52,0)*+{\bigoplus_{p}V^{xy}_{p}\otimes V^{pz}_{u}}="1",
(-32,20)*+{\bigoplus_{p}V^{yx}_{p}\otimes V^{pz}_{u}}="2",
(32,20)*+{\bigoplus_{q}V^{yq}_{u}\otimes V^{xz}_{q}}="3",
(52,0)*+{\bigoplus_{q}V^{yq}_{u}\otimes V^{zx}_{q}}="4",
(-32,-20)*+{\bigoplus_{r}V^{xr}_{u}\otimes V^{yz}_{r}}="5",
(32,-20)*+{\bigoplus_{r}V^{yz}_{r}\otimes  V^{rx}_{u}}="6",
\ar^-(0.4){\bigoplus_{p}R^{xy}_{p}\otimes id_{V^{pz}_{u}}} "1";"2"
\ar^-{F^{yxz}_{u}} "2";"3"
\ar^-(0.6){\bigoplus_{q} id_{V^{yq}_{u}}\otimes R^{xz}_{q}} "3";"4"
\ar^-(0.6){F^{xyz}_{u}} "1";"5"
\ar^-{\bigoplus_{r}(id_{V^{yz}_{r}}\otimes R^{xr}_{u})\,\cdot\,
\mathrm{swap}} "5";"6"
\ar^-(0.4){F^{yzx}_{u}} "6";"4"
\end{xy}
\bigskip
\end{equation*}
\begin{equation*}
\begin{xy}
(-52,0)*+{\bigoplus_{p}V^{xy}_{p}\otimes V^{pz}_{u}}="1",
(-32,20)*+{\bigoplus_{p}V^{yx}_{p}\otimes V^{pz}_{u}}="2",
(32,20)*+{\bigoplus_{q}V^{yq}_{u}\otimes V^{xz}_{q}}="3",
(52,0)*+{\bigoplus_{q}V^{yq}_{u}\otimes V^{zx}_{q}}="4",
(-32,-20)*+{\bigoplus_{r}V^{xr}_{u}\otimes V^{yz}_{r}}="5",
(32,-20)*+{\bigoplus_{r}V^{yz}_{r}\otimes  V^{rx}_{u}}="6",
\ar^-(0.4){\bigoplus_{p}(R^{yx}_{p})^{-1}\otimes id_{V^{pz}_{u}}} "1";"2"
\ar^-{F^{yxz}_{u}} "2";"3"
\ar^-(0.6){\bigoplus_{q} id_{V^{yq}_{u}}\otimes (R^{zx}_{q})^{-1}} "3";"4"
\ar^-(0.6){F^{xyz}_{u}} "1";"5"
\ar^-{\bigoplus_{r}( id_{V^{yz}_{r}}\otimes (R^{rx}_{u})^{-1})\,\cdot\,
\mathrm{swap}} "5";"6"
\ar^-(0.4){F^{yzx}_{u}} "6";"4"
\end{xy}
\end{equation*}
\caption{Hexagonal equations with fusion spaces.}\label{fig:hex1}
\end{figure}
\begin{rem}
A fusion category with strict unitors and  trivial triangle equations results in some simplifying rules which reduce the number of pentagonal and hexagonal equations. For example, in the Fibonacci model instead of solving 32 equations, one needs to only solve one equation. Every fusion matrix has four indices, $F_{abc}^{d}$, when one of these indices is the vacuum anyon, I, the triangle equation is trivially satisfied and $F$ is identity. Hence, only one $F$ whose all indices are $\tau$ is non-identity. For further information, See \cite{mythesis}. 
\end{rem}
\begin{rem}
Note that fusion spaces are hom-sets of a $URFC$. Hence, they come equipped with a well-defined inner product, which transforms them into Hilbert spaces. Given the definition above, this category is also closed under the tensor product, moreover, the direct sum of the $F$ and $R$ matrices are well-behaved. 
\end{rem}
\section{The 2-Category of Topological Quantum Computation}
In the 2-categorical framework, we assign:
\begin{itemize}
    \item \textbf{0-morphisms} to anyon types,
    \item \textbf{1-morphisms} to vector spaces of 1-morphisms between tensor products of anyons (i.e., the vector spaces include creation and annihilation operators),
    \item \textbf{2-morphisms} to fusion and braiding matrices, which act as isomorphisms between these vector spaces.
\end{itemize}

We \emph{skeletalise} the 2-category at the level of 0-morphisms, since in physical systems, isomorphic or equivalent anyon types are considered indistinguishable.

The hom-categories between anyon types (i.e., between 0-morphisms) contain fusion spaces \(V_{ab}^c\) as simple objects, and fusion and braiding matrices as isomorphisms between the corresponding vector spaces.

The \emph{pentagon equation} serves as a coherence condition in this setting. Assuming the pentagonator \(\pi\) is the identity (i.e., working in a strict 2-category), the associativity constraints reduce to commutative diagrams among vector spaces, Figure~\ref{fig:penta2}. 

To construct the 2-categorical picture of topological quantum computation, let us first restrict our attention to one of the widely known models, i.e. Fibonacci anyons.

\subsection{The 2-Category of Fibonacci Anyons}
The Fibonacci model is the simplest yet universal model of topological quantum computation. The model has two anyon types, $L= \{1, \tau\}$, where $1$ is the trivial or vacuum anyon, and $\tau$ is the Fibonacci anyon. The only non-trivial fusion rule is, 
$$\tau \times \tau = 1 + \tau$$
The non-zero fusion spaces are $\mathcal{V} = \{V_{\tau \tau}^1, V_{\tau \tau}^\tau, V_{1\tau}^\tau, V_{\tau1}^\tau, V_{11}^1\}$, where all of them are 1-dimensional vector spaces, $V \cong \mathbb C$. As we indicated before, we skeletalise the 2-category at the 0-morphism level. Hence, any combination or tensor product of Fibonacci anyons, is isomorphic either to $1$ or $\tau$. Each isomorphism between anyonic combinations belongs to a vector space, $V$,  which is an object of Hom-categories,  $\mathrm{HomCat}_{\tau}$ or $\mathrm{HomCat}_{1}$. $\mathrm{HomCat}$'s are fusion categories whose objects are fusion spaces. Note that $V_{1\tau}^\tau$ or $V_{\tau1}^\tau$ are the variations of $V_{\tau\tau}^1$. 
\begin{align*}
& \mathrm{HomCat}_\tau = \langle V_{1\tau}^\tau, V_{\tau \tau}^\tau \rangle, 
&  \mathrm{HomCat}_1= \langle V_{11}^1, V_{\tau \tau}^1\rangle
\end{align*}
So a combination of $(\tau \times \tau) \times \tau$ introduces Fusion space, $V_{(\tau\tau)\tau}^\tau$, which belongs to $\mathrm{HomCat}_{\tau}$ and has an expansion based on simple fusion spaces as follows, 
\[
V_{(\tau\tau)\tau}^\tau \cong V_{\tau\tau}^{\textcolor{red}{1}}\otimes V_{\textcolor{red}{1}\tau}^\tau \oplus V_{\tau\tau}^{\textcolor{red}{\tau}} \otimes V_{{\textcolor{red}{\tau}}\tau}^\tau.
\]
Grouping anyons differently results in another yet isomorphic fusion space with an expansion as follows:
\[
V_{\tau(\tau\tau)}^\tau \cong V_{\tau{\textcolor{red}{\tau}}}^\tau \otimes V_{\tau\tau}^{\textcolor{red}{\tau}} \oplus V_{\tau {\textcolor{red}{1}}}^\tau \otimes V_{\tau\tau}^{\textcolor{red}{1}}.
\]
Fusion matrices are isomorphisms between these vector spaces. \[
F_{\tau \tau \tau} ^\tau: V_{(\tau\tau)\tau}^\tau \longrightarrow V_{\tau(\tau\tau)}^\tau
\]
Note that if we consider another vector space such as, $V_{(11)1}^1$ and decompose this vector space based on primary fusion spaces, we will see it belongs to $\mathrm{HomCat_1}$. As fusion spaces, $V_{11}^1$ belong to $\mathrm{HomCat_1}$. This space, therefore, does not have any decomposition based on fusion spaces in $\mathrm{HomCat_\tau}$. 
\[
V_{(11)1}^1 = V_{11}^1 \otimes V_{11}^1
\]
We can clearly observe the superselection rule here. Assume we have a state $\ket{\psi} \in  V_{11}^1$ and another state $\ket{\phi} \in V_{\tau\tau}^\tau$, these two fusion spaces belong to different Hom-categories,  are not able to form a coherent state as $\ket{\phi}+\ket{\psi}$. 

The next step is to recover the pentagonal equation from the equations of braided monoidal 2-categories. First, note that the version of the pentagonal equation we intend to recover is the equation based on isomorphic vector spaces Figure~\ref{fig:pentag}, and it is expressed in Appendix E of Kiatev's paper \cite{kitaev2006}.  To recover the equation, we need to consider the definition of the pentagonator, and modification $\pi$ in monoidal 2-categories Figure \ref{fig:pentagonator}. 
\begin{figure}[h]
    \centering
    \includegraphics[scale=0.7]{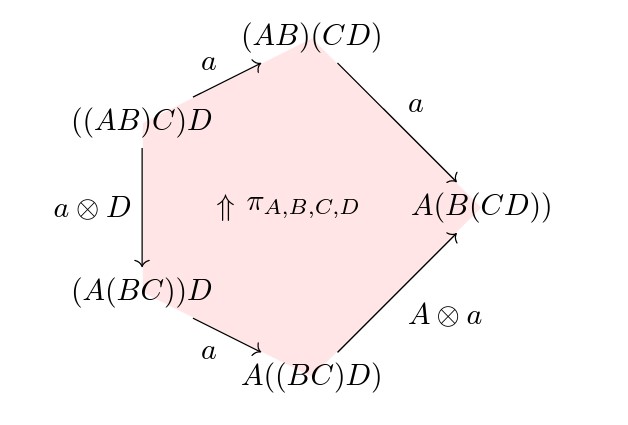}
    \caption{Pentagonator $\pi$ in monoidal 2-categories.} \label{fig:pentagonator}
    \label{fig:pentag}
\end{figure}
The monoidal 2-category is strict. Thus, all $a$-morphisms and modification $\pi$ are identities. The tensor products $\tau$'s on the vertices is equivalent to $\tau$ as the 2-category is skeletal at the level of objects. Each non-trivial arrow represents the collection of operations between two objects. 
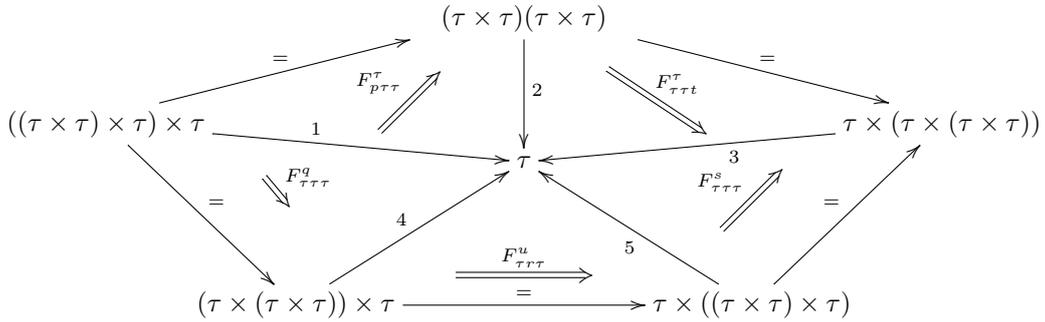
\begin{figure}[!ht]
\begin{equation*}
\begin{xy}
(-55,0)*+{((\tau \times \tau)\times \tau)\times \tau}="1",
(0,14)*+{(\tau \times \tau)(\tau \times \tau)}="2",
(55,0)*+{\tau \times (\tau \times (\tau \times \tau))}="3",
(-30,-24)*+{(\tau \times (\tau \times \tau))\times \tau}="4",
(30,-24)*+{\tau \times ((\tau \times \tau) \times \tau)}="5",
(-0, -5)*+{\tau}="6"; 
(-20, -2)*+{}="7"; 
(-10, 8)*+{}="8"; 
(10, 8)*+{}="10"; 
(25, -2)*+{}="9";
(-35, -6)*+{}="11"; 
(-30, -12)*+{}="12"; 
(-10, -20)*+{}="13"; 
(10, -20)*+{}="14"; 
(25, -15)*+{}="15"; 
(35, -5)*+{}="16";

\ar^-{=} (-4,0)+"1"; (-4,0)+"2"
\ar^-{=} (4,0)+"2"; (4,0)+"3"
\ar^{=} "1";"4"
\ar^{=} "4";"5"
\ar^{=} "5";"3"
\ar^{1} "1";"6"
\ar^{2} "2";"6"
\ar^{3} "3";"6"
\ar^{4} "4";"6"
\ar^{5} "5";"6"
\ar@{=>}^{F_{p\tau\tau}^\tau} "7";"8"
\ar@{=>}^{F_{\tau\tau t}^\tau} "10";"9"
\ar@{=>}^{F_{\tau\tau \tau}^q} "11";"12"
\ar@{=>}^{F_{\tau r \tau}^u} "13";"14"
\ar@{=>}^{F_{\tau\tau \tau}^s} "15";"16"
\end{xy}
\end{equation*}
\caption{Recovering pentagonal equation from monoidal 2-category. }\label{fig:pentag2}
\end{figure}
\begin{align*}
    & 1 = V_{((\tau\tau)\tau)\tau}^\tau, 2 = V_{(\tau\tau)(\tau\tau)}^\tau, 3 = V_{\tau(\tau(\tau\tau))}^\tau, 4 = V_{(\tau(\tau\tau))\tau}^\tau, 5 = V_{\tau((\tau\tau)\tau)}^\tau
\end{align*}
We can furthermore represent the composition of 2-morphisms in a more explicit form as follows:
\begin{center}
\begin{tikzpicture}[>=latex, thick]

% Nodes
\node (A) at (0,0) {$\tau$};
\node (B) at (5,0) {$\tau$};

% Curved arrows upwards (left to right)
\draw[->] (A) to[bend left=80] node[above] {1} (B);
\draw[->] (A) to[bend left=50] node[above] {2} (B);
\draw[->] (A) to[bend left=20] node[above] {3} (B);

% Curved arrows downwards (left to right)
\draw[->] (A) to[bend right=20] node[below] {4} (B);
\draw[->] (A) to[bend right=50] node[below] {5} (B);
\draw[->] (A) to[bend right=80] node[below] {1} (B);
% Double vertical arrows between curved arrow groups
\draw[double,->] (2.2,1.5) -- (2.2,1.3);
\draw[double,->] (2.2,1.2) -- (2.2,0.6);
\draw[double,->] (2.2,0.4) -- (2.2,-0.3);
\draw[double, ->] (2.2, -0.5)--(2.2, -0.9); 
\draw[double, ->] (2.2, -1.3)--(2.2, -1.7); 
\end{tikzpicture}
\end{center}
The vertical composition of filling 2-morphisms or Fusion matrices, $F$, results in the Pentagonal equation~\ref{fig:penta2}. The right hand side is the filling 2-morphism of pentagonator, $\pi$, which is identity. Since the background 2-category we are considering is a strict monoidal 2-category. 
\begin{equation}
    (F_{\tau\tau\tau}^s)^{-1} \odot (F_{\tau r\tau}^u)^{-1} \odot (F_{\tau \tau \tau }^q)^{-1} \odot (F_{\tau \tau t}^u) \odot (F_{p\tau \tau }^u)= \pi 
\end{equation}
One can recover the hexagonal equations, Figure~\ref{fig:hex1} from hexagonators. 
\begin{figure}
\begin{equation*}
\begin{tikzcd}[row sep = large, column sep = scriptsize]
    &
    BAC
    \arrow[r]
    \arrow[dd, phantom, pos=0.1, "\scriptstyle{\xRightarrow{(R'_{(A|B,C)})^{-1}}}"]
    \arrow[ddr, phantom, "\scriptstyle{ \xRightarrow{R_{(A,R_{B,C})}^{-1}}}"]
    &
    BCA
    \arrow[dr, start anchor = east]
    \arrow[dd, phantom, pos=0.9, "\scriptstyle{ \xRightarrow{R'_{(A|B,C)}}}"]
    \\
    ABC
    \arrow[ur, end anchor = west]
    \arrow[dr, end anchor = west]
    \arrow[urr, start anchor = east, end anchor = south]
    &&&
    CBA
    \\
    &
    ACB
    \arrow[r]
    \arrow[urr, start anchor = north, end anchor = west]
    &
    CAB
    \arrow[ur, start anchor = east]
\end{tikzcd}
\qquad
\begin{tikzcd}[row sep = large, column sep = scriptsize]
    &
    BAC
    \arrow[r]
    \arrow[drr, start anchor = south, end anchor = west]
    &
    BCA
    \arrow[dr, start anchor = east]
    \\
    ABC
    \arrow[ur, end anchor = west]
    \arrow[dr, end anchor = west]
    \arrow[drr, start anchor = east, end anchor = north]
    &&&
    CBA
    \\
    &
    ACB
    \arrow[uu, phantom, pos=0.1, "\scriptstyle{ \xRightarrow{(R'_{(A,B|C)})^{-1}}}"]
    \arrow[uur, phantom, "\scriptstyle{ \xRightarrow{R_{(R_{A,B},C)}}}"]
    \arrow[r]
     &
    CAB
    \arrow[ur, start anchor = east]
    \arrow[uu, phantom, pos=0.9, "\scriptstyle{ \xRightarrow{R'_{(A,B|C)}}}"]
\end{tikzcd}
\end{equation*}
\caption{Hexagonators in braided monoidal 2-categories.}\label{hexagonators}
\end{figure}
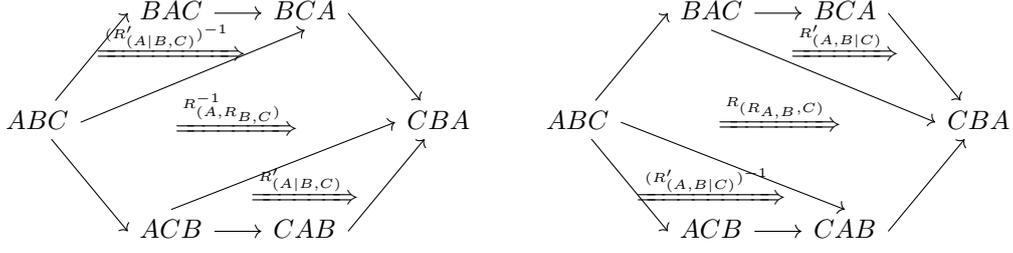
Consider two hexagonators Figure \ref{hexagonators}, and similarly, assume that all side 1-morphisms and filling 2-morphisms are identity. Each combination of the tensor product of $\tau$'s is isomorphic to $\tau$. Thus, each arrow represents the vector spaces of operations between them. The vector spaces are isomorphic, and the filling 2-morphisms will be a combination of braiding matrices, $R$.  The Fibonacci example provides an intuition for us to formulate the general case.  
\begin{equation}\label{hex3}
\begin{xy}
(-52,0)*+{(\tau\times \tau)\times \tau}="1",
(-32,20)*+{(\tau \times \tau )\times \tau}="2",
(32,20)*+{\tau \times (\tau \times \tau)}="3",
(52,0)*+{\tau \times (\tau \times \tau)}="4",
(-32,-20)*+{\tau \times (\tau\times \tau)}="5",
(32,-20)*+{\tau \times (\tau\times \tau)}="6",
(0, 0)*+{\tau}="7",
(-30, 4)*+{}="8",
(-28, 10)*+{}="9",
(-1, 10)*+{}="10",
(5, 10)*+{}="11",
(22, 10)*+{}="12",
(28, 7)*+{}="13",
(25, -5)*+{}="15",
(32, -1)*+{}="14",
(5, -10)*+{}="16",
(-1, -10)*+{}="17",
(-35, -5)*+{}="18",
(-32, -9)*+{}="19",

\ar^-(0.4){=} "1";"2",
\ar^-(0.4){=} "2";"3",
\ar^-(0.6){=} "3";"4"
\ar^-(0.6){=} "1";"5"
\ar^-(0.4){=} "5";"6"
\ar^-(0.4){=} "6";"4"
\ar@{=>}^{R_{\tau\tau}^\tau \times id} "8";"9"
\ar@{=>}^{F_{(\tau\tau)\tau}^\tau} "10";"11"
\ar@{=>}^{F_{(\tau\tau)\tau}^\tau} "12";"13"
\ar@{=>}^{id \times R_{\tau\tau}^\tau} "14";"15"
\ar@{=>}^{id \times R_{\tau\tau}^\tau} "17";"16"
\ar@{=>}^{F_{(\tau\tau)\tau}^\tau} "18";"19"

\ar^-{V_{(\tau\tau)\tau}^\tau} "1";"7"
\ar^-{V_{(\tau\tau)\tau}^\tau} "2";"7"
\ar^-{V_{\tau(\tau\tau)}^\tau} "3";"7"
\ar^-{V_{\tau(\tau\tau)}^\tau} "4";"7"
\ar^-{V_{\tau(\tau\tau)}^\tau} "5";"7"
\ar^-{V_{\tau(\tau\tau)}^\tau} "6";"7"
\end{xy}
\bigskip
\end{equation}

\subsection{The 2-Category of Ising Anyons}
Ising anyons or Majorana fermions are currently the main candidate for topological quantum computing as the proposals for realising them are more straightforward to implement. However, this model is not universal, and they can only implement Clifford gates.  The model includes three anyons, $L = \{ 1, \sigma, \psi\}$ and the fusion rules are as follows:
\begin{align*}
    & \sigma \times \sigma = 1 + \psi,  & \sigma \times \psi = \psi \times \sigma = \sigma \\
    & 1 \times \sigma = \sigma \times 1 & 1 \times \psi = \psi \times 1 = \psi 
\end{align*}
These fusion rules result in the following 1-dimensional fusion spaces, as $V \cong \mathbb{C}$. 
\begin{equation*}
\mathcal{V} = \{V_{\sigma\sigma}^1, V_{\sigma\sigma}^\psi, V_{\sigma\psi}^\sigma, V_{\sigma1}^\sigma, V_{\psi\psi}^1,  V_{\psi\sigma}^\sigma, V_{1\sigma}^\sigma, V_{11}^1\}    
\end{equation*}
Given anyon types and knowing their fusion rules and spaces, we can construct the $\mathrm{HomCat}$ as follows:  
\begin{align*}
    & \mathrm{HomCat}_{\sigma} = <V_{\sigma1}^\sigma, V_{\psi\sigma}^\sigma, V_{1\sigma}^\sigma>, & \mathrm{HomCat}_{\psi} = <V_{\sigma\sigma}^\psi, V_{\psi1}^\psi>,\\
    & \mathrm{HomCat}_1= <V_{\sigma\sigma}^1, V_{\psi\psi}^1, V_{11}^1>
\end{align*}
The Hom-category we use for quantum computing is $\mathrm{HomCat}_{\sigma}$ since it includes fusion spaces whose tensor products provide a 2-dimensional vector space suitable for encoding qubits. The computational space is based on three Ising anyons, given below
\[
(\sigma \times \sigma ) \times \sigma = 2\sigma 
\]
Hence, the 2-dimensional space constructed from fusion spaces of $\mathrm{HomCat}_\sigma$ is as follows: 
\[ 
V_{(\sigma\sigma)\sigma}^\sigma = V_{\sigma\sigma}^{\textcolor{red}{\psi}} \otimes  V_{\textcolor{red}{\psi}\sigma}^\sigma \oplus V_{\sigma\sigma}^{\textcolor{red}{1}} \otimes  V_{\textcolor{red}{1}\sigma}^\sigma
\]
Similar to Fibonacci anyons, one can solve pentagonal and hexagonal equations for Ising model given the above decomposition and find solutions correspondingly. 
\subsection{The 2-Category of Fermionic Moore-Read}
The example of Fermionic Moore-Read is more complex than Fibonacci anyons as the non-abelian anyon $\sigma$ has a different anti-particle, $\sigma'$ and the label set has more anyons. The label set is,
\[
L = \{ 1,\alpha,\alpha',\sigma,\sigma', \psi \}
\]
If the subset $\{1, \alpha, \psi, \alpha'\}$ is identified with $\mathbb{Z}_4 =\{0, 1, 2, 3\} $, then the fusion rules for this subset agrees with $\mathbb{Z}_4$. The rest of the fusion rules is summarized in the following; 
\begin{align*}
& \sigma \times \sigma' = 1+\psi  & \sigma \times \sigma = \sigma' \times \sigma'= \alpha + \alpha' \\
& \sigma \times \psi = \psi \times \sigma = \sigma & \sigma' \times \psi = \psi \times \sigma' = \sigma' \\
& \sigma \times \alpha = \sigma \times \alpha'=  \alpha' \times \sigma = \alpha \times \sigma = \sigma' \\
& \sigma' \times \alpha = \sigma' \times \alpha'=  \alpha' \times \sigma' = \alpha \times \sigma' = \sigma
\end{align*}
These fusion rules result in constructing fusion categories over simple objects. 
\begin{align}
    & \mathrm{HomCat}_{\sigma} = <V_{\sigma1}^\sigma, V_{\sigma'\alpha}^\sigma, V_{\sigma\psi}^\sigma, V_{\sigma'\alpha'}^\sigma >, & \mathrm{HomCat}_\alpha = <V_{1\alpha}^{\alpha}, V_{\sigma\sigma}^\alpha. V_{\sigma'\sigma'}^\alpha, V_{\alpha'\psi}^\alpha> \\
    & \mathrm{HomCat}_{\alpha'} = <V_{1\alpha'}^{\alpha'}, V_{\sigma'\sigma'}^\alpha, V_{\sigma\sigma}^{\alpha'}, V_{\alpha\psi}^{\alpha'}>, &\mathrm{HomCat}_{\sigma'} = <V_{\sigma'1}^{\sigma'}, V_{\sigma\alpha}^{\sigma'}, V_{\sigma'\psi}^{\sigma'}, V_{\sigma\alpha'}^{\sigma'}>\\
    & \mathrm{HomCat}_{\psi} = <V_{\sigma\sigma'}^\psi, V_{\psi1}^\psi, V_{\alpha\alpha}^\psi, V_{\alpha'\alpha'}^{\psi}>, & \mathrm{HomCat}_1=<V_{11}^1, V_{\alpha\alpha'}^1, V_{\psi\psi}^1, V_{\sigma\sigma'}^1>
\end{align}
For instance, if we have three $\sigma$ anyons, their fusion results in $\sigma'$ with multiplicity 2 and this can be further expanded in terms of vector spaces.   
\[
(\sigma \times \sigma)\times \sigma = 2\sigma'
\]
The vector space, $V_{(\sigma\sigma)\sigma}^{\sigma'}$, has a decomposition based on fusion spaces in $\mathrm{HomCat}_{\sigma'}$. Given the vector spaces in $\mathrm{HomCat}_{\sigma'}$ and their decomposition, we can further obtain solutions of pentagonal and hexagonal equations. 
\[
V_{(\sigma\sigma)\sigma}^{\sigma'} = V_{\sigma\sigma}^{\textcolor{red}{\alpha}} \otimes V_{\textcolor{red}{\alpha}\sigma}^{\sigma'} \oplus V_{\sigma\sigma}^{\textcolor{red}{\alpha'}} \otimes V_{\textcolor{red}{\alpha'}\sigma}^{\sigma'}
\]
\subsection{TQC Category as a Braided Fusion 2-Category}
Having the above explicit examples, we can now define the general case of the 2-category of topological quantum computing. A definition for fusion 2-categories was proposed in~\cite{douglas2018fusion}. As one might expect, each categorical component of fusion categories is generalized to 2-categories. However, we do not spell out all parallel definitions, as it would be cumbersome. A deep understanding of the notions defined for fusion categories illuminates the 2-categorical correspondence. Moreover, the version we work with does not invoke all peculiarities of fusion 2-categories.   

\begin{definition}
A \emph{fusion 2-category} is a finite semisimple monoidal 2-category that has left and right duals for objects and a simple monoidal unit.
\end{definition}

This definition appears to be the 2-categorical analogue of a fusion category. A few points, however, need to be clarified:
\begin{itemize}
    \item Unlike the 1-categorical case, where morphisms between non-isomorphic simple objects are zero morphisms, this condition is not generally satisfied in 2-categories. In our setting, we assume that 1-morphisms between non-isomorphic simple objects are zero morphisms, mimicking semisimplicity at the object level.
    \item Fusion 2-categories are defined to be semisimple monoidal 2-categories, indicating that their Hom-categories are semisimple linear categories. Furthermore, we work with 2-categories whose Hom-categories are fusion categories.
\end{itemize}

The initial background 2-category is taken to be a strict braided fusion 2-category. For our purposes, we assume all coherency conditions at the level of 0-morphisms are trivial. That is, at the level of objects in the 2-category, associativity 1-morphisms are identity 1-morphisms. We also assume strictness and skeletality at the level of objects. Whether this holds in general for arbitrary fusion 2-categories remains to be shown in future work. Nonetheless, we conjecture that, unlike in monoidal categories, skeletality and strictness should coexist in this higher-categorical setting.

We now define a braided fusion 2-category. A definition for braided monoidal 2-categories, inspired by Kapranov and Voevodsky, was proposed by Baez and Neuchl in~\cite{baez1996higher}. Their setting is that of a semi-strict 2-category, which suffices for our purposes. Fusion 2-categories are defined in~\cite{douglas2018fusion}. Therefore, our desired structure is a synthesis of these two frameworks: we incorporate Baez and Neuchl's braiding into Douglas and Reutter's definition of a fusion 2-category. Additionally, we require that all Hom-categories are unitary braided fusion categories, enabling concrete computations within them.

\begin{definition}
A \textbf{braided fusion 2-category} is a 2-category satisfying:
\begin{itemize}
    \item Objects are simple.
    \item Each object has equivalent left and right duals.
    \item A tensor product, $\times $ and sum  $+$ exists between objects.
    \item A braiding, as defined in~\cite{baez1996higher}, is present.
    \item Hom-categories are unitary braided fusion categories. 
\end{itemize}
\end{definition}

This 2-category serves as the foundational structure for the 2-category of topological quantum computing. At the level of objects, we work with the hardware of topological quantum computation. Anyonic types and their fusion and braiding behavior are described at this level. In particular, we use the symbols $\times$ and $+$ to indicate fusion between anyons and the sum over different outcomes of anyonic interactions.

At the level of 1-morphisms and 2-morphisms, we formalize the mathematical structure of models of quantum computation. In the Hom-categories, 1-morphisms correspond to fusion spaces arising from the fusion of anyons, while 2-morphisms correspond to fusion and braiding matrices, derived from solving the pentagonal and hexagonal equations.

\begin{definition}
The \textbf{2-category of topological quantum computing} is a braided fusion 2-category such that:
\begin{itemize}
    \item Simple objects are anyonic types, denoted $X_i$.
    \item The tensor product $\times$ between objects specifies fusion rules:
    \[
    X_i \times X_j = X_k + \dots + X_l.
    \]
    \item At the level of objects, associativity is strict.
    \item For each object $X_i$, there exists a Hom-category, denoted as $\mathrm{HomCat}_{X_i}$.
    \item Objects in $\mathrm{HomCat}_{X_i}$ (i.e., 1-morphisms) are fusion spaces indexed by anyonic types, $V_{ij}^k$.
    \item Other fusion spaces are constructed via tensor product and direct sum:
    \[
    V_{ijk}^l \cong \bigoplus_{p} V_{ij}^p \otimes V_{pk}^l.
    \]
    \item 2-morphisms (morphisms in $\mathrm{HomCat}_{X_i}$) are fusion and braiding matrices.
    \item Fusion matrices $F_{ijk}^l$ in each Hom-category satisfy the pentagonal equation (cf. Figure~\ref{fig:penta2}).
    \item Braiding matrices $R_{ij}^k$ in each Hom-category satisfy the hexagonal equations (cf. Figure~\ref{fig:hex1}).
\end{itemize}
\end{definition}

\section{Superselection rules}
Superselection rules prevent forming coherent states between eigenstates of certain observables. The notion was introduced in 1952 by Wick (1909--1992), Wightman, and Wigner (1902--1995) to impose further restrictions on quantum theory beyond selection rules~\cite{giulini2016superselection}. The rule is particularly mentioned by Kitaev as an axiom in the introduction of his paper~\cite{kitaev2006}. In the context of bosonic and fermionic states, this rule is clear because the direct sum of bosonic and fermionic Hilbert spaces imposes such a restriction. However, in the categorical description of anyonic states, one needs to pay extra attention to avoid forming non-coherent states. 

In the 2-categorical version, however, this rule is baked into the structure. Let us first review the precise definition of superselection rules, then see how it behaves in anyonic theories, and finally explain how our 2-categorical framework incorporates the rule automatically.

\begin{definition}
Two quantum states, $\ket{\psi_1}$ and $\ket{\psi_2}$, are separated by a \textbf{superselection rule} if 
\[
\bra{\psi_1} A \ket{\psi_2} = 0
\]
for all physical observables $A$. Hence, no observable can create coherent superpositions between them, and they are indistinguishable from a classical mixture of the two states.
\end{definition}

Applying this rule to topological quantum computing prevents the creation of coherent superpositions of anyonic states with different total charges. Consider the Moore--Read anyonic theory, where the fusion of two non-abelian anyons $\sigma$ results in two anyonic types, $\alpha$ and $\alpha'$, 
\[
\sigma \times \sigma = \alpha + \alpha'.
\]
In the categorical picture, one might naïvely assume that one can form a superposition such as
\[
\ket{\psi_{\sigma\sigma}^\alpha} + \ket{\psi_{\sigma\sigma}^{\alpha'}}.
\]
However, the total charge of these two states differs, and thus the sum does not correspond to a physically coherent quantum state~\cite{kitaev2006}.

Our approach handles superselection rules automatically because the computational categories are now the Hom-categories labeled by objects of the 2-category. Hence, when computation is restricted to a Hom-category, for example $\mathrm{Hom}_X$, one is working with vector spaces indexed only by the single object $X$, as explicitly shown in the above examples. For instance, $\ket{\psi_{\sigma\sigma}^\alpha} \in V_{\sigma\sigma}^\alpha$ and $\ket{\psi_{\sigma\sigma}^{\alpha'}} \in V_{\sigma\sigma}^{\alpha'}$, and these vector spaces belong to two different Hom-categories. 

\section{Acknowledgements}
I would like to thank Steve Simon and Aleks Kissinger, as well as Urs Schreiber for valuable discussions. I also thank Antoine (Jack) Jacquier and acknowledge support from UKRI under Grant Ref: EP/W032643/1.

\bibliographystyle{alpha}
\bibliography{sample}

\newcommand{\etalchar}[1]{$^{#1}$}
\begin{thebibliography}{AAA{\etalchar{+}}25}

\bibitem[AAA{\etalchar{+}}25]{aasen2025roadmap}
David Aasen, Morteza Aghaee, Zulfi Alam, Mariusz Andrzejczuk, Andrey Antipov, Mikhail Astafev, Lukas Avilovas, Amin Barzegar, Bela Bauer, Jonathan Becker, et~al.
\newblock Roadmap to fault tolerant quantum computation using topological qubit arrays.
\newblock {\em arXiv preprint arXiv:2502.12252}, 2025.

\bibitem[Ahm20]{ahmadi2020monoidal}
Fatimah Ahmadi.
\newblock Monoidal 2-categories: A review.
\newblock {\em arXiv preprint arXiv:2011.02830}, 2020.

\bibitem[Ahm22]{mythesis}
Fatimah~Rita Ahmadi.
\newblock Bicategorical aspects of topological quantum computation, 2022.

\bibitem[AK22]{ahmadi2022top}
Fatimah~Rita Ahmadi and Aleks Kissinger.
\newblock Topological quantum computation through the lens of categorical quantum mechanics.
\newblock {\em arXiv preprint arXiv:2211.03855}, 2022.

\bibitem[Ali12]{alicea2012new}
Jason Alicea.
\newblock New directions in the pursuit of majorana fermions in solid state systems.
\newblock {\em Reports on progress in physics}, 75(7):076501, 2012.

\bibitem[Ati88]{atiyah1988}
Michael~F Atiyah.
\newblock Topological quantum field theory.
\newblock {\em Publications Math{\'e}matiques de l'IH{\'E}S}, 68:175--186, 1988.

\bibitem[BN96]{baez1996higher}
John~C Baez and Martin Neuchl.
\newblock Higher dimensional algebra: I. braided monoidal 2-categories.
\newblock {\em Advances in Mathematics}, 121(2):196--244, 1996.

\bibitem[DR18]{douglas2018fusion}
Christopher~L Douglas and David~J Reutter.
\newblock Fusion 2-categories and a state-sum invariant for 4-manifolds.
\newblock {\em arXiv preprint arXiv:1812.11933}, 2018.

\bibitem[FLW02]{freedman2002}
Michael~H Freedman, Michael Larsen, and Zhenghan Wang.
\newblock A modular functor which is universal{\P} for quantum computation.
\newblock {\em Communications in Mathematical Physics}, 227(3):605--622, 2002.

\bibitem[Giu16]{giulini2016superselection}
Domenico Giulini.
\newblock Superselection rules.
\newblock {\em From chemistry to consciousness: The legacy of Hans Primas}, pages 45--70, 2016.

\bibitem[Kit03]{kitaev2003}
A~Yu Kitaev.
\newblock Fault-tolerant quantum computation by anyons.
\newblock {\em Annals of Physics}, 303(1):2--30, 2003.

\bibitem[Kit06]{kitaev2006}
Alexei Kitaev.
\newblock Anyons in an exactly solved model and beyond.
\newblock {\em Annals of Physics}, 321(1):2--111, 2006.

\bibitem[Lei98]{leinster1998basic}
Tom Leinster.
\newblock Basic bicategories.
\newblock {\em arXiv preprint math/9810017}, 1998.

\bibitem[Lei14]{leinster2014}
Tom Leinster.
\newblock {\em Basic category theory}, volume 143.
\newblock Cambridge University Press, 2014.

\bibitem[RW18]{rowell2018mathematics}
Eric Rowell and Zhenghan Wang.
\newblock Mathematics of topological quantum computing.
\newblock {\em Bulletin of the American Mathematical Society}, 55(2):183--238, 2018.

\bibitem[SFN06]{sarma2006}
Sankar~Das Sarma, Michael Freedman, and Chetan Nayak.
\newblock Topological quantum computation.
\newblock {\em Physics today}, 59(7):32--38, 2006.

\bibitem[Sim23]{simon2023topological}
Steven~H Simon.
\newblock {\em Topological quantum}.
\newblock Oxford University Press, 2023.

\bibitem[Wan10]{wang2010}
Zhenghan Wang.
\newblock {\em Topological quantum computation}.
\newblock Number 112. American Mathematical Soc., 2010.

\end{thebibliography}
\end{document}